\documentclass[showpacs,preprintnumbers,twocolumn,aps,twocolumn,nobibnotes,nofootinbib,citeautoscript]{revtex4-1}

\usepackage{amsmath,amssymb,times,graphicx,bm,bbold}
\usepackage[usenames,dvipsnames,svgnames,table]{xcolor}
\usepackage[colorlinks=true, urlcolor=Blue, linkcolor=Blue, citecolor=Blue, pdftex]{hyperref}

\def\<{\langle}
\def\>{\rangle}
\DeclareMathOperator{\Tr}{Tr}
\newcommand{\ve}[1]{\boldsymbol{#1}}
\def\smallapprox{\,{\approx}\,}

\newcommand{\JSTAT}{J.~Stat.~Mech.}

\newcommand{\PRL}{Phys.~Rev.~Lett.}

\newcommand{\JPCS}{J.~Phys.: Conf.~Ser.}

\newcommand{\JPSJ}{J.~Phys.~Soc.~Jap.}

\newcommand{\RMP}{Rev.~Mod.~Phys}

\begin{document}

\title{Mutual information in heavy-fermion systems}

\author{\firstname{Francesco} \surname{Parisen Toldin}}
\email{francesco.parisentoldin@physik.uni-wuerzburg.de}
\author{\firstname{Toshihiro} \surname{Sato}}
\email{Toshihiro.Sato@physik.uni-wuerzburg.de}
\author{\firstname{Fakher F.} \surname{Assaad}}
\email{assaad@physik.uni-wuerzburg.de}
\affiliation{\mbox{Institut f\"ur Theoretische Physik und Astrophysik, Universit\"at W\"urzburg, Am Hubland, 97074 W\"urzburg, Germany}}

\begin{abstract}
A key notion in  heavy-fermion systems is the entanglement between conduction electrons and localized spin degrees of freedom. To study these systems from this point of view, we compute the mutual information in a ferromagnetic and antiferromagnetic Kondo lattice model in the presence of geometrical frustration.   Here the interplay between the Kondo effect,  the Ruderman-Kittel-Kasuya-Yosida interaction,  and geometrical frustration  leads  to  partial Kondo screened, conventional Kondo insulating, and antiferromagnetic phases.    In each of these states the mutual information  follows an area law, the coefficient of which shows sharp crossovers (on our finite lattices) across phase transitions.    Deep in the respective phases,  the area law coefficient can be understood in terms of  simple direct product wave functions thereby  yielding an accurate measure of the entanglement in each phase. The above-mentioned results stem from approximation-free auxiliary field quantum Monte Carlo simulations. 
\end{abstract}

\maketitle

\section{Introduction}
\label{sec:intro}
The Kondo effect describes  the screening of a spin-1/2 magnetic impurity  embedded in a metallic environment~\cite{Hewson}.
At high temperatures the spin degree of freedom is decoupled from the conduction electrons  and below the Kondo scale a many body entangled state of the spin and conduction electrons  emerges.  
To  quantify  entanglement between   a bipartition $A$ and $B$ of a system,  one traces out the degrees of freedom $B$  to obtain a reduced density matrix, $\hat{\rho}_{A} = \text{Tr}_{\cal{H}_B} \hat{\rho}$, the  Renyi entropy of which, $S_n(A) = (\ln  \text{Tr} \hat{\rho}_{A}^n) / (1-n)   $,  corresponds to the entanglement entropy. 
Taking  one subsystem to be the  impurity spin, and the other the conduction electrons, the Kondo effect  can be elegantly characterized  by the  transfer of  $\ln(2)$  thermal entropy at high  temperatures to $\ln(2)$  entanglement entropy in the ground state, as recently computed in a Kondo impurity model \cite{WCPI-18} and in a spin-$1/2$ chain sharing the same low-energy behavior \cite{Sorensen07}. The energy scale  at which this transfer from the thermal entropy to the entanglement entropy occurs is the Kondo temperature. 
More generally, local entanglement is an important feature of two-level dissipative systems \cite{KLH-07,LeHur-08}.

In the presence of a lattice of spins  Kondo coupled to conduction electrons, corresponding to heavy-fermion systems \cite{Coleman07_rev},  the above picture breaks down.  In fact spins can now interact through the indirect Ruderman-Kittel-Kasuya-Yosida (RKKY) exchange interaction~\cite{RKKY},  and thereby compete with Kondo screening.  Comparing these two energy scales, it becomes apparent that  Kondo screening   dominates  when the exchange interaction between the localized spins and the conduction electrons, $J_{\text{K}}$, is positive and large, and that the RKKY interaction  dominates at small values of  $J_{\text{K}}$.  The intricate interplay between these two effects on nonfrustrated lattices leads to a quantum phase transition (QPT) between  disordered  and   ordered magnetic phases~\cite{Doniach77,Assaad99a,LRVW-07,SPNYGYK-14}. The Kondo effect can be switched off by considering  $J_{\text{K}}<0$, thereby promoting magnetically ordered phases~\cite{Assaad-KLQMC}.
In addition,
 geometrical frustration is found to be of experimental relevance in many heavy-fermion materials such as CePdAl, Pr${}_2$Ir${}_2$O${}_7$, YbAgGe, YbAl${}_3$C${}_3$, Yb${}_2$Pt${}_2$Pb \cite{akito16,nakatsuji06, kim_frustkondo08,sengupta_frustkondo10,kato_frustkondo08}, where quantum phases do not easily fit into the aforementioned cases.
 Geometrical frustration can lead to  so-called partial Kondo screened (PKS)  phases where  frustration is alleviated by selective spatial  screening localized spins \cite{MNYU-10,NYPK-13,Aulbach2015,PK-17,SAG-18}.
 The essence of all  aforementioned states can be captured by direct product variational wave functions  from which one can directly assess the degree of entanglement between the spins and conduction electrons.
 Entanglement entropies, although not presently experimentally accessible in heavy-fermion systems,
 lend themselves to an experimental measure in systems of cold atoms \cite{IMPTLRM-15}, which, in turn, allow one to realize Kondo lattice models \cite{GHGXJYZDLR-10}. Alternatively, entanglement properties have been recently proposed to be experimentally studied by engineering the so-called entanglement Hamiltonian in cold atom systems \cite{DVZ-18}. In this context, Ref.~\cite{PTA-18} introduces a numerically exact method to determine the entanglement Hamiltonian in interacting models of fermions.

In this paper we investigate a Kondo lattice model Hamiltonian amenable to negative-sign-free quantum Monte Carlo (QMC) simulations that  provide specific realizations of the states discussed above. Using recently developed methods  to compute the Renyi entropies \cite{Grover13} with the auxiliary field QMC,  we compute the  mutual information and show that, deep in the respective phases, the numerical value of the area law coefficient can be well understood  in terms of the product state wave function description of the phase supplemented by fluctuations, if necessary.   Furthermore,  we observe a singular  behavior of the area law coefficient across phase transitions. 

\section{Model}
\label{sec:model}
We consider the generalized Kondo lattice model on the honeycomb lattice
introduced in \cite{SAG-18} with Hamiltonian
\begin{equation}
    \hat{H} = -t\sum_{ \< \ve{ i},\ve{j} \> }   \hat{\ve{c}}^{\dagger}_{\ve{i}} \hat{\ve{c}}^{\phantom\dagger} _{\ve{j}}+
    J_{\text{K}}\sum_{\ve{i}}   \frac{1}{2} \hat{\ve{c}}^{\dagger}_{\ve{i}}\ve{\sigma}\hat{\pmb{c}}^{\phantom\dagger}_{\ve{i}}\cdot \hat{\ve{S}}_{\ve{i}}
    +J_{z} \sum_{ \< \<  \ve{ i},\ve{j}  \> \> }\hat{S}^z_{\ve{i}} \hat{S}^z_{\ve{j}}
\label{KondoH}
\end{equation}
where the first sum extends over the nearest-neighbor sites and describes the hopping of conduction electrons, 
$\hat{\ve{c}}^{\dagger}_{\ve{i}} = \left( \hat{c}^{\dagger}_{\ve{i}, \uparrow}, \hat{c}^{\dagger}_{\ve{i}, \downarrow} \right)  $, giving rise to  the well-known semimetallic band dispersion \cite{CNGPNG-09}, the second sum accounts for the Kondo screening interaction between conduction electrons and spin-1/2 local moments $\hat{\boldsymbol{S}}_{\ve{i}}$, while the third term is the next-nearest-neighbor antiferromagnetic interaction between localized spins  encodes frustration effects.
This model can be solved without encountering the negative sign problem \cite{SAG-18};
here and in the following we consider half-filling for the conduction electron and use $t = 1$ as the energy unit.
 Due to the antiferromagnetic coupling $J_z$
this half-filled Kondo lattice model on the honeycomb lattice with geometric frustration exhibits PKS phases alongside the conventional Kondo insulator (KI) and antiferromagnetically ordered phases \cite{SAG-18}.

 The mutual information  $I_n(\Gamma_c, \Gamma_S)$ between two subsystems of conduction electrons $\Gamma_{c}$ and of localized spins $\Gamma_{S}$ is
 \begin{equation}
   I_n(\Gamma_c, \Gamma_S) \equiv S_n(\Gamma_{c}) + S_n(\Gamma_{S}) - S_n(\Gamma_{c} \cup \Gamma_{S}),
   \label{mutual_def}
 \end{equation}
 where $S_n(\Gamma)$ is the $n$th Renyi entropy for a subsystem $\Gamma$.
 Here we take $\Gamma_{c}$ as a compact subset of $N$ conduction electron sites, and $\Gamma_{S}$ as the corresponding $N$ localized spin sites coupled to the subset $\Gamma_{c}$.
 Assuming the ubiquitous area law for the entanglement entropy \cite{ECP-10}, $I_n(\Gamma_c, \Gamma_S)$ results are proportional to the size of the boundary shared between $\Gamma_{c}$ and $\Gamma_{S}$:
 \begin{equation}
   I_n(\Gamma_c, \Gamma_S) \simeq \alpha 2N.
   \label{arealaw}
 \end{equation}
 The mutual information is also defined in terms of the von Neumann entanglement entropy, corresponding to the limit $n\rightarrow 1$ in Eq.~(\ref{mutual_def}). In this case $I_1(\Gamma_c, \Gamma_S)$ satisfies  \cite{OhyaPetzBook}
 \begin{equation}
   I_1(\Gamma_c, \Gamma_S) \ge \frac{\<O_c O_S\>-\<O_c\>\<O_S\>}{{\parallel} O_c{\parallel}^2{\parallel} O_S{\parallel}^2},
   \label{mutual_bound}
 \end{equation}
 where the numerator represents the connected correlation of two arbitrary operators $O_c$ and $O_S$ acting on the subsystem $\Gamma_c$ and $\Gamma_S$, respectively, and ${\parallel} X{\parallel}\equiv \{\text{sup} \sqrt{\<\psi|X^\dag X|\psi\>}, \<\psi|\psi\>=1\}$ is the norm of an operator $X$. According to Eq.~(\ref{mutual_bound}), $I_1(\Gamma_c,\Gamma_S)$ bounds all mutual correlations of operators in $\Gamma_c$ and $\Gamma_S$, thus providing an operator-independent entanglement measure.   Due to the above bound, $I_1(\Gamma_c,\Gamma_S)$   captures both high- and low-energy  scales. 
 
 Here, we shall consider the mutual information for Renyi index $n=2$.
 The coefficient $\alpha$ introduced in Eq.~(\ref{arealaw}) is the main quantity investigated in this work.
 In the presence of Kondo screening $I_n(\Gamma_c, \Gamma_S)$ essentially counts the number of Kondo  singlets formed between $\Gamma_c$ and $\Gamma_S$, such that
 in the limit $J_{\text{K}} \rightarrow \infty$  Eq.~(\ref{arealaw}) holds exactly with $\alpha = \ln(2)$. 

\section{Method}
\label{sec:method}
We have investigated the Hamiltonian of Eq.~(\ref{KondoH}) by means of auxiliary field QMC \cite{Blankenbecler81,White89,AF_notes} simulations, using the method of Ref.~\cite{SAG-18} which, in essence, consists in a fermion representation of localized spins obtained via Lagrange multipliers. We refer to Ref.~\cite{SAG-18} for more details on the formulation. A similar technique can be used to simulate the canonical ensemble \cite{WAPT-17}. Simulations have been performed using the ALF package \cite{ALF}.
To compute the Renyi entropies we have used a method introduced in Ref.~\cite{Grover13}, and also used in \cite{ALPT-13,CSS-14,DP-15,DP-16}, which allows to
formulate the reduced density matrix within auxiliary field QMC.
Beside Renyi entropies, the technique can be exploited to unbiasedly determine the entanglement Hamiltonian \cite{PTA-18}.
Reference \cite{PTA-18b} provides a short review of computational approaches to entanglement in interacting fermionic systems.

\begin{figure}[b]
  \centering
    \includegraphics[width=0.8\linewidth]{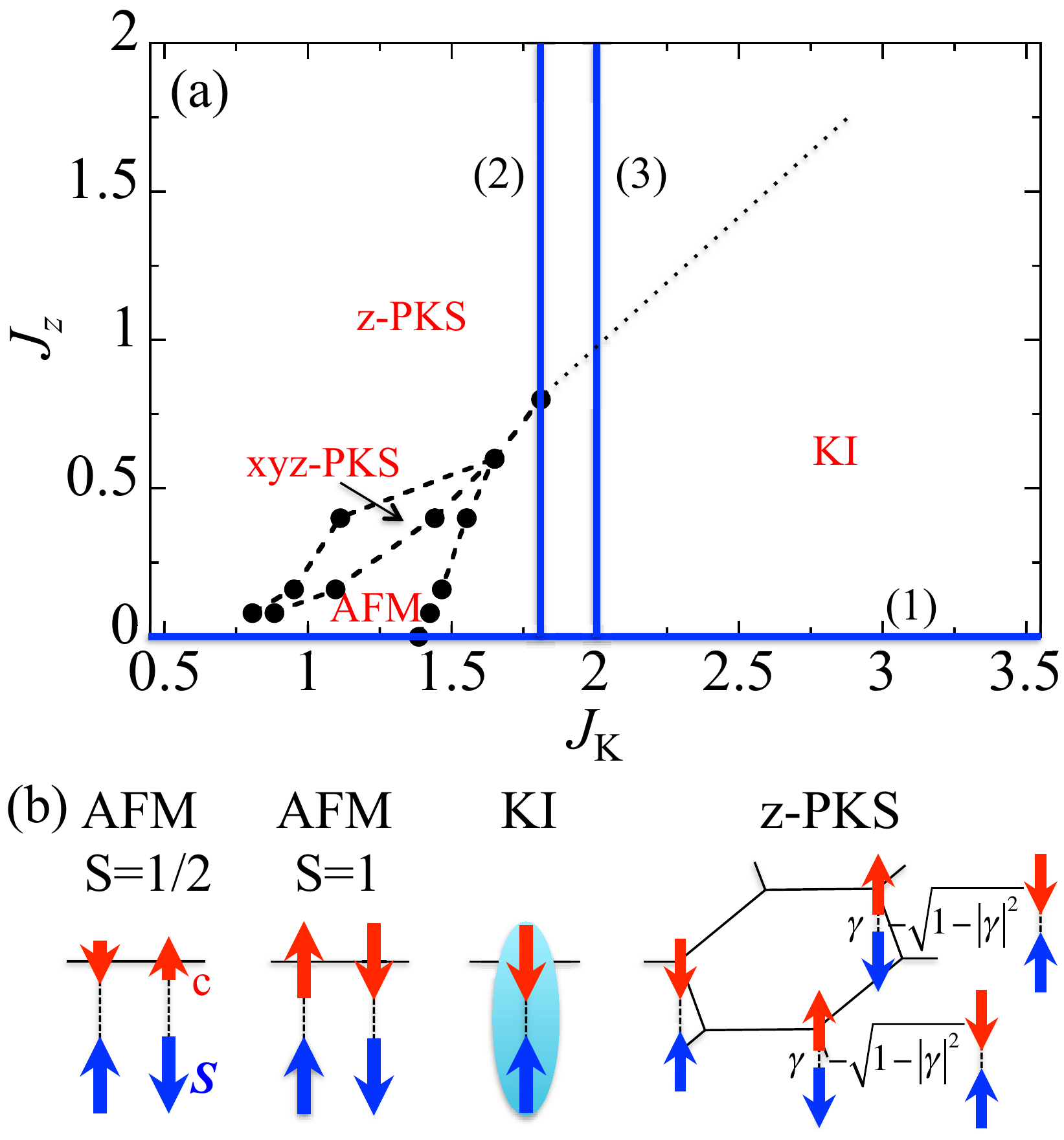}
    \caption{(a) Ground-state phase diagram with antiferromagnetic (AFM), out-of-plane PKS ($z$-PKS), spin-rotation symmetry breaking PKS ($xyz$-PKS), and KI phases from QMC simulations~\cite{SAG-18}. Dashed lines connects transition points and the dotted line sketches the expected boundary between the KI and $z$-PKS phases. The three thick lines indicate the scans of the phase diagram considered here. (b) Mean-field schematic picture of $S=1/2$ and $S=1$ AFM, KI, and $z$-PKS phases.}
    \label{phase_diagram}
\end{figure}

\begin{figure*}
  \centering
  \includegraphics[width=\linewidth]{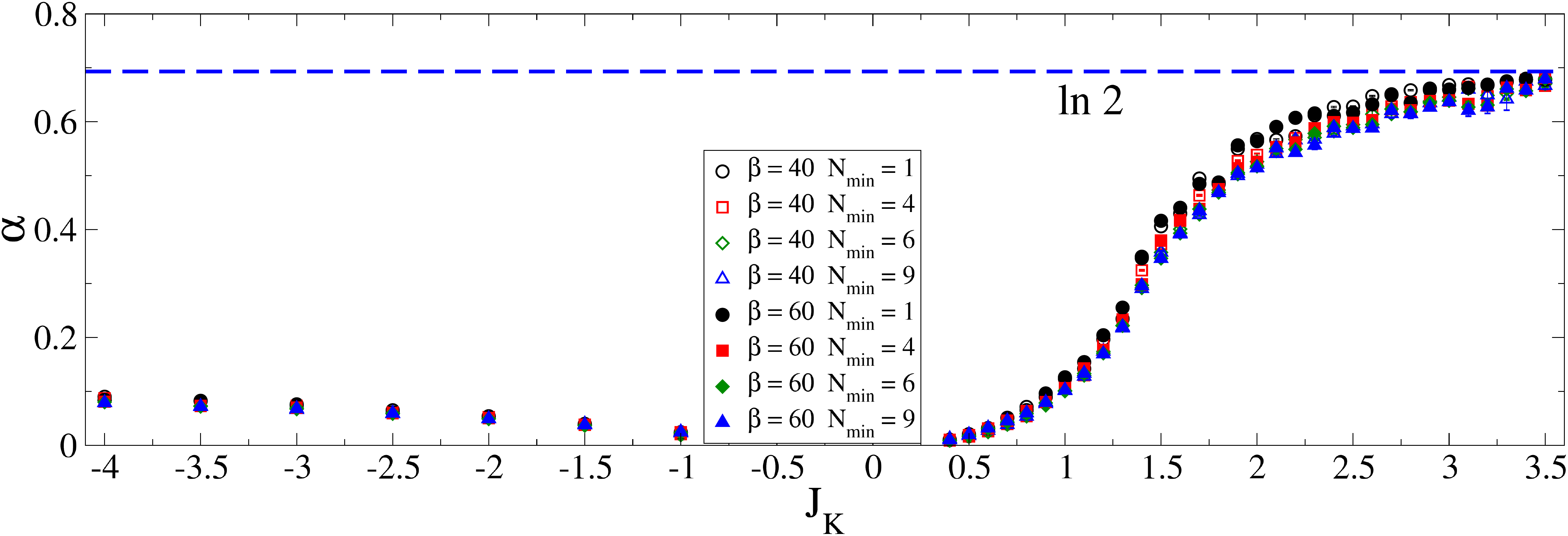}
  \caption{Coefficient $\alpha$ of the area law for the mutual information along the line (1) on Fig.~\ref{phase_diagram}(a) and also for $J_{\text{K}}\le-1$, $J_z=0$. We consider two inverse temperatures $\beta=40$, $60$, and for four values of the minimum subsystem size taken into account in the fits.}
  \label{scan1}
\end{figure*}

\section{Results}
\label{sec:results}
In Fig.~\ref{phase_diagram}(a) we reproduce the rich phase diagram of the model for $J_{\text{K}}>0$, $J_z \ge 0$.  At finite $J_z$ the model has a reduced U(1) spin symmetry corresponding to spin rotations around the $z$ axis, as well as the point group and translation symmetries of the honeycomb  lattice.  The KI phase breaks no symmetries, the in-plane antiferromagnetic ($xy$-AFM) phase  breaks the U(1) spin symmetry, and  the $z$-PKS phase breaks nematically the point group  and has reduced translation symmetry.   The $xyz$-PKS phase differs  from the $z$-PKS one in that it additionally breaks the U(1) spin symmetry \cite{SAG-18}. 
In Fig.~\ref{phase_diagram}(a) we also show three lines on the phase diagram where we analyze the mutual information.
Moreover, we study the entanglement for $J_{\text{K}}<0$, $J_z=0$, where the model favors the formation of an effective spin $S=1$ Heisenberg antiferromagnet in the strong-coupling limit.
In all the QMC data presented here we have simulated a lattice $6\times 6$ unit cells  corresponding  to 144  orbitals. 

We first consider a scan for $J_z=0$, where the model reduces to a standard Kondo lattice model on the honeycomb lattice with the conventional RKKY driven AFM phase and KI phase. A fit of $I_2(\Gamma_c, \Gamma_S)$ for up to nine choices of $\Gamma_c$, $\Gamma_S$ to the right-hand side of Eq.~(\ref{arealaw}) 
allows us to extract the coefficient $\alpha$ shown in Fig.~\ref{scan1}, for two inverse temperatures $\beta=40$, $60$, and as a function of the minimum subsystem size $N_{\text{min}}$ taken into account in the fits.
More technical details on the chosen subsystems $\Gamma_c$, $\Gamma_S$ are reported in Appendix \ref{sec:mutual}.
We observe consistent results for the fitted values of $\alpha$.
In  the KI phase conduction and localized electrons are paired into a spin singlet, such that for $J_{\text{K}}\gg 0$ the ground state approaches a product of single-site singlet wave functions shown in Fig.~\ref{phase_diagram}(b), giving a $\ln(2)$ entanglement entropy per pair.
Indeed, in the KI phase $\alpha$ reaches an asymptotic value $\ln(2)$ for $J_{\text{K}}\gtrsim 3$. Interestingly, a change of concavity in the plot of $\alpha$ occurs around the QPT between the AFM and KI phases, at $J_{\text{K}}\simeq 1.4$ \cite{SAG-18}.
In the AFM phase  at $J_{\text{K}}  > 0$   the conduction electron local moment aligns antiparallel to the  spin and is reduced in magnitude due to charge fluctuations.  
The essential features of the AFM phase can be captured by a product wave function, in which the  total  local moment of conduction electrons and spins  form a N\'eel order  as depicted in Fig.~\ref{phase_diagram}(b).   Entanglement between  spins and conduction electrons originates from subleading Kondo screening  as argued in Ref.~\cite{Assaad-KLQMC}  and also from   long-wavelength spin-wave fluctuations of the  N\'eel   order parameter. 
In Fig.~\ref{scan1} we also show $\alpha$ for a ferromagnetic coupling $J_{\text{K}}<0$. Different than the $J_{\text{K}}>0$ case, here the Kondo coupling favors the formation of a spin triplet in the ground state of the model, without Kondo screening.
Starting from the limit $J_{\text{K}}\rightarrow-\infty$, where on each lattice site the spin singlet state $|S = 0\>$ is projected away, a finite large value of $J_{\text{K}}<0$ gives rise to an antiferromagnetic exchange term between $S = 1$ states on the honeycomb lattice. Thus, in this situation
the system reduces
to a Heisenberg $S = 1$ model. Its antiferromagnetic ground state is well captured by the semiclassical large-$S$ expansion, and consists in a N\'eel state, illustrated in Fig.~\ref{phase_diagram}(b). Since such a ground state is primarily built on $|S=1, S_z=\pm 1\>$ states, the entanglement between conduction electrons and localized spins is expected to be small. This observation is clearly reflected in the $\alpha$ coefficient shown in Fig.~\ref{scan1}, whose values for $J_{\text{K}}<0$ are substantially smaller than for $J_{\text{K}}>0$. For instance, we find $\alpha\simeq 0.1$ for $J_{\text{K}}=1$ and $\alpha\simeq 0.024$ for $J_{\text{K}}=-1$.
The computed value of $\alpha$ grows on reducing $J_{\text{K}}<0$, but appears to saturate to a value significantly smaller than the limiting value $\ln(2)$
found for $J_{\text{K}}\rightarrow \infty$.

\begin{figure}
  \centering
  \includegraphics[width=0.85\linewidth]{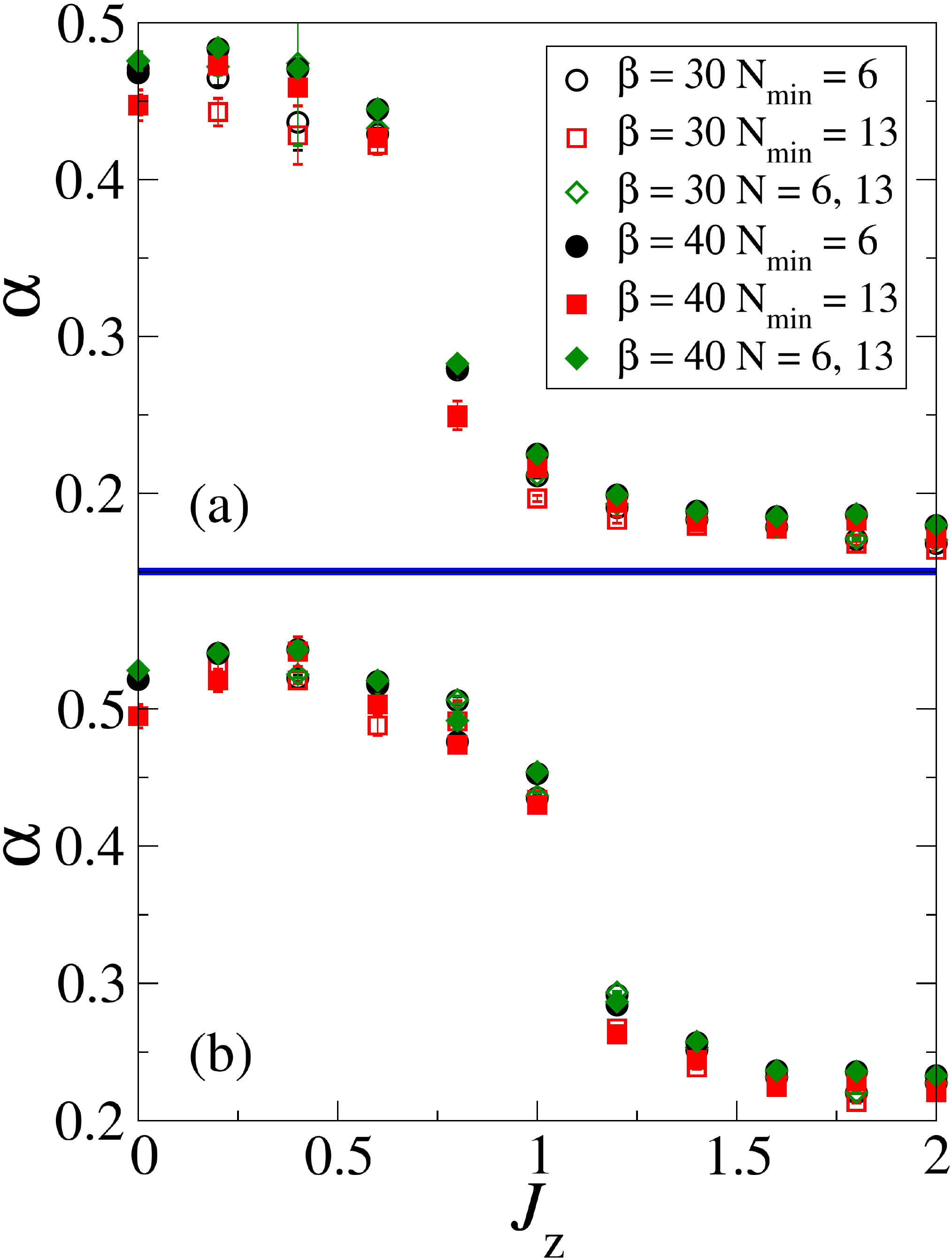}
  \caption{Coefficient $\alpha$ of the area law for the mutual information
    along line (2) (above) and line (3) (below) on Fig.~\ref{phase_diagram}(a), for inverse temperatures $\beta=30$, $40$, and three different fits (see main text).}
  \label{scan23}
\end{figure}

In order to investigate the $z$-PKS phase we compute the mutual information as a function of $J_z$ for fixed $J_{\text{K}}=1.8$. As illustrated in Fig.~\ref{phase_diagram}(a), this second scan crosses the conventional KI and $z$-PKS phases. Due to an enlarged unit cell expected in the $z$-PKS phase \cite{SAG-18}, we have in this case considered only three possible subsystems $\Gamma_c$, $\Gamma_S$, with equal size $N=6$, $13$, $22$. In view of the limited amount of available data, and  in order to reliably study the coefficient $\alpha$ we have considered three possible linear fits: a fit including all data, a fit disregarding the smallest size $N=6$, and a fit disregarding the largest size $N=22$. In Fig.~\ref{scan23}(a) we show the corresponding results, for two inverse temperatures $\beta=30$, $40$. Despite fluctuations larger than the error bars, indicating sizable corrections to Eq.~(\ref{arealaw}), we observe a clear trend in $\alpha$, which decreases from $\alpha\smallapprox 0.45$ to $\alpha\smallapprox 0.17$.
Moreover, the curve shows again a change in the curvature at a value of $J_z$ approximately consistent with the onset of a QPT between the KI and $z$-PKS phases, located at $J_z\smallapprox 0.8$ \cite{SAG-18} [see also Fig.~\ref{phase_diagram}(a)]. For $J_z\gtrsim 1.4$, $\alpha$ saturates to a plateau $\alpha\smallapprox 0.17$. Such a value, which approaches the limit $J_z \rightarrow \infty$ at fixed $J_{\text{K}}$ is in fact $J_{\text{K}}$-dependent, as shown by the computation of $\alpha$ along path (3) in Fig.~\ref{phase_diagram}(a).
In Fig.~\ref{scan23}(b) we show the resulting $\alpha$, whose curves are qualitatively similar to the case of Fig.~\ref{scan23}(a), but saturate to a large value $\alpha\smallapprox 0.23$ for large $J_z$.

The emergence of an area law with a coefficient $\alpha < \ln(2)$ confirms the mechanism of partial Kondo screening in the $z$-PKS phase, emerging from the competition of the antiferromagnetic interaction and the Kondo coupling.
To understand the structure of the ground state we consider the limit $J_{\text{K}}$, $J_z\rightarrow\infty$, i.e., the atomic limit $t=0$ in which the honeycomb lattice decomposes into two independent triangular sublattices. In this limit the Hamiltonian (\ref{KondoH}) commutes with the $z$ component of the total spin operator on site $\ve{i}$, $S^{\text{tot},z}_{\ve{i}} \equiv S^{\text{c},z}_{\ve{i}} + S^{z}_{\ve{i}}$, with $S^{\text{c},z}_{\ve{i}}\equiv (1/2)\hat{\pmb{c}}^{\dagger}_{\ve{i}} \sigma_z\hat{\pmb{c}}^{\phantom\dagger}_{\ve{i}}$. Accordingly, one expects the ground state to have $S^{\text{tot},z}_{\ve{i}}=0$, such that its wave function is constructed from the states $|+\>_{\ve{i}}\equiv |\uparrow, \downarrow\>_{\ve{i}}$ and $|-\>_{\ve{i}}\equiv | \downarrow, \uparrow\>_{\ve{i}}$, whereas states obtained with the remaining base vectors $|\uparrow, \uparrow\>_{\ve{i}}$ and $|\downarrow, \downarrow\>_{\ve{i}}$ are gapped. In the Hilbert space spanned by $\{|+\>_{\ve{i}}, |-\>_{\ve{i}}\}$ the Hamiltonian (\ref{KondoH}) is, up to a constant,
\begin{equation}
  \hat{H} = \frac{J_z}{4}\sum_{\<\< \ve{i},\ve{j} \> \>} \hat{Z}_{\ve{i}} \hat{Z}_{\ve{j}} + \frac{J_{\text{K}}}{2} \sum_{\ve{i}} \hat{X}_{\ve{i}} + \ldots,
  \label{HIsing}
\end{equation}
where the operators $\hat{Z}_{\ve{i}}$, $\hat{X}_{\ve{i}}$ are defined by $\hat{Z}_{\ve{i}} |\pm\>_{\ve{i}}=\pm |\pm\>_{\ve{i}}$, $\hat{X}_{\ve{i}}|\pm\>_{\ve{i}}= |\mp \>_{\ve{i}}$ and satisfy the commutation rule of the SU(2) algebra. Therefore, in the atomic limit and close to the ground state the Hamiltonian (\ref{KondoH}) reduces to that of an antiferromagnetic transverse-field Ising model, on a triangular lattice. 
Beyond the atomic limit,   the effective low-temperature Hamiltonian acquires additional interactions which we have already anticipated in Eq.~(\ref{HIsing}).  However since  for a finite, but large, $J_{\text{K}}/t$ , and still  $J_{\text{K}}\ll J_z$ the additional states are gapped, we expect the ground state to be well  representable  in the  $\{|+\>_{\ve{i}}, |-\>_{\ve{i}}\}$   basis. 
It is known that the above Ising model has a three sublattice structure  and that the  sign of the sixfold clock term in the Landau-Ginzburg functional determines if the   ground state will be 
hierarchical,    $  |\Psi\>  =  |+\>_{\ve{i}} |-\>_{\ve{j}}  (|+\>_{\ve{k}}+|-\>_{\ve{k}})/\sqrt{2} $,  or uniform  \cite{Blankschtein84,Moessner01b},
\begin{equation}
  \begin{split}
    |\Psi\> = &|+\>_{\ve{i}} \left(\gamma |+\>_{\ve{j}}  - \sqrt{1-|\gamma|^2}|-\>_{\ve{j}} \right) \\
    &\otimes\left(\gamma |+\>_{\ve{k}}  - \sqrt{1-|\gamma|^2}|-\>_{\ve{k}}\right).
  \end{split}
  \label{gs}
\end{equation}
Here,  $\ve{i}$, $\ve{j}$, $\ve{k}$ label the  three sites of the sublattice structure.   Numerical calculations in  Ref.~\cite{SAG-18}  point to the uniform ground state, so we
use this product wave function to  account for our  mutual information results. 
Here, $|\gamma|\le 1$ controls the partial polarization of the sites $\ve{j}$ and $\ve{k}$ and   the direct product wave function is illustrated in 
Fig.~\ref{phase_diagram}(b).
As we discuss in Appendix \ref{sec:renyi}, for the state of Eq.~(\ref{gs}), $\alpha=-(2/3)\ln (1 - 2|\gamma|^2 + 2|\gamma|^4)$, which takes values between $0$ and $2\ln(2)/3\simeq 0.46$, thus including the plateaus found in Fig.~\ref{scan23}.

\begin{figure*}
  \centering
  \includegraphics[width=0.9\linewidth]{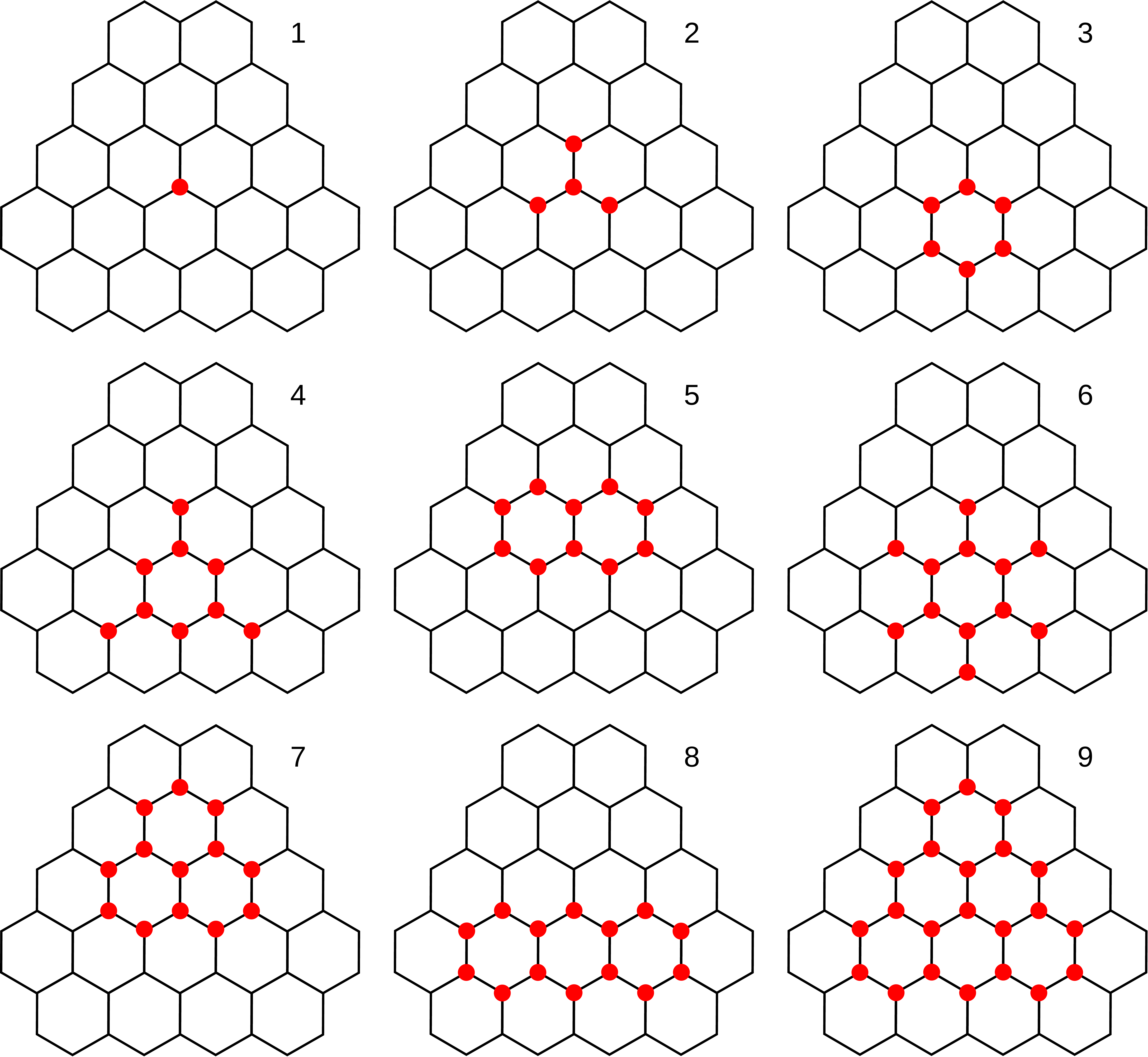}
  \caption{Partition of lattice sites used to compute the mutual information $I_2(\Gamma_c, \Gamma_S)$. For each illustrated subset of sites, $\Gamma_c$ is the corresponding set of conduction electron sites, and $\Gamma_S$ the set of localized spin sites coupled to $\Gamma_c$. When $J_z>0$, due to the enlarged unit cell we have computed $I_2(\Gamma_c, \Gamma_S)$ using only subsystems no. $3$, no. $7$, and no. $9$.}
  \label{all_subsystems}
\end{figure*}

\section{Conclusions and Discussion}
\label{sec:conclusions}
The bound on the  von Neumann  entanglement entropy based mutual information  presented in Eq.~(\ref{mutual_bound}) implies that for our choice  of $\Gamma_c$ and $\Gamma_S$  the mutual information  picks up IR as well as UV physics.    Deep in a phase,  or in a sink in the renormalization-group parlance, where the correlation length is finite one can model the ground state with a direct  product wave function that provides a modeling of all energy scales.  With this wave function  the coefficient of the area law of the mutual information can be computed.    In the Kondo lattice model considered here, we have  an explicit realization of  three phases,  AFM, PKS, and KI.  The simple direct product wave functions for these three phases  presented in Fig.~\ref{phase_diagram}(b) have an area  law coefficient set by $\alpha = 0 $ for the N\'eel   representation of the AFM phase,  $ \alpha =  \ln(2) $ for the strong-coupling KI phase, and   $\alpha$ bounded by  $2\ln(2)/3$ for the 
PKS phase.  The KI value of $\alpha$ is exact in the strong $J_{\text{K}}$ limit  and is very well reproduced  when $J_{\text{K}}$  is comparable to half the bandwidth, $W=6$.     Deep in the PKS phase  $\alpha$  shows a $J_{\text{K}}$ dependent plateau  upon enhancing $J_z$.  The value of this plateau depends on the degree of partial Kondo screening as  described by $\gamma$ in Eq.~(\ref{gs}). 
  Clearly, the N\'eel wave function  shows no entanglement, but of course does not capture  fluctuations  leading to Goldstone modes and to entanglement.  For these phases, both  for positive and negative values of $J_{\text{K}}$ the  mutual information remains {\it small} but does not vanish.   It is however noticeable that $\alpha$ is larger  in the AFM phase at $J_{\text{K}} > 0$ than at $J_{\text{K}} < 0$. We understand this difference in terms of subleading Kondo screening present for  the antiferromagnetic model  but absent for the ferromagnetic one.    Computations  of the single-particle spectral  function  \cite{Assaad-KLQMC} confirm this point of view. 
  
On our finite lattices we have observed clear crossovers  in the area law coefficient of the mutual information across phase transitions.    How this behavior  reflects the  criticality of the transition as well as possible corrections to the area  law at critical points is left to future investigations. 

 From the technical point of view, the present calculation of the mutual information  does not lead to a noticeable increase in computational effort.   It can be used as a standard {\it observable independent }   measure  to pick up QPTs and thereby map out  phase diagrams in various  heavy-fermion systems and beyond.   This   is very similar to  unsupervised machine learning algorithms  aiming at automatically mapping out phase diagrams \cite{Broecker17}.  As mentioned above it also provides further information  on phases.  In this context we note that this quantity has recently been used \cite{Hofmann18} to validate the understanding of a Kondo breakdown transition.   

\begin{acknowledgments}
We would like to thank T.  Grover and J. S. Hofmann for insightful  discussions.
F.P.T. thanks the German Research Foundation (DFG)
through Grant No. AS120/13-1 of the FOR 1807.
T.S. thanks the DFG for financial support from Grant No. AS120/14-1.
F.F.A. thanks the DFG through SFB 1170 ToCoTronics.
We gratefully acknowledge the Gauss Centre for Supercomputing (GCS) under the project pr53ju for allocation of CPU time on the SuperMUC computer at the Leibniz Supercomputing Center.
\end{acknowledgments}

\appendix

\begin{figure}
  \centering
  \includegraphics[width=0.85\linewidth]{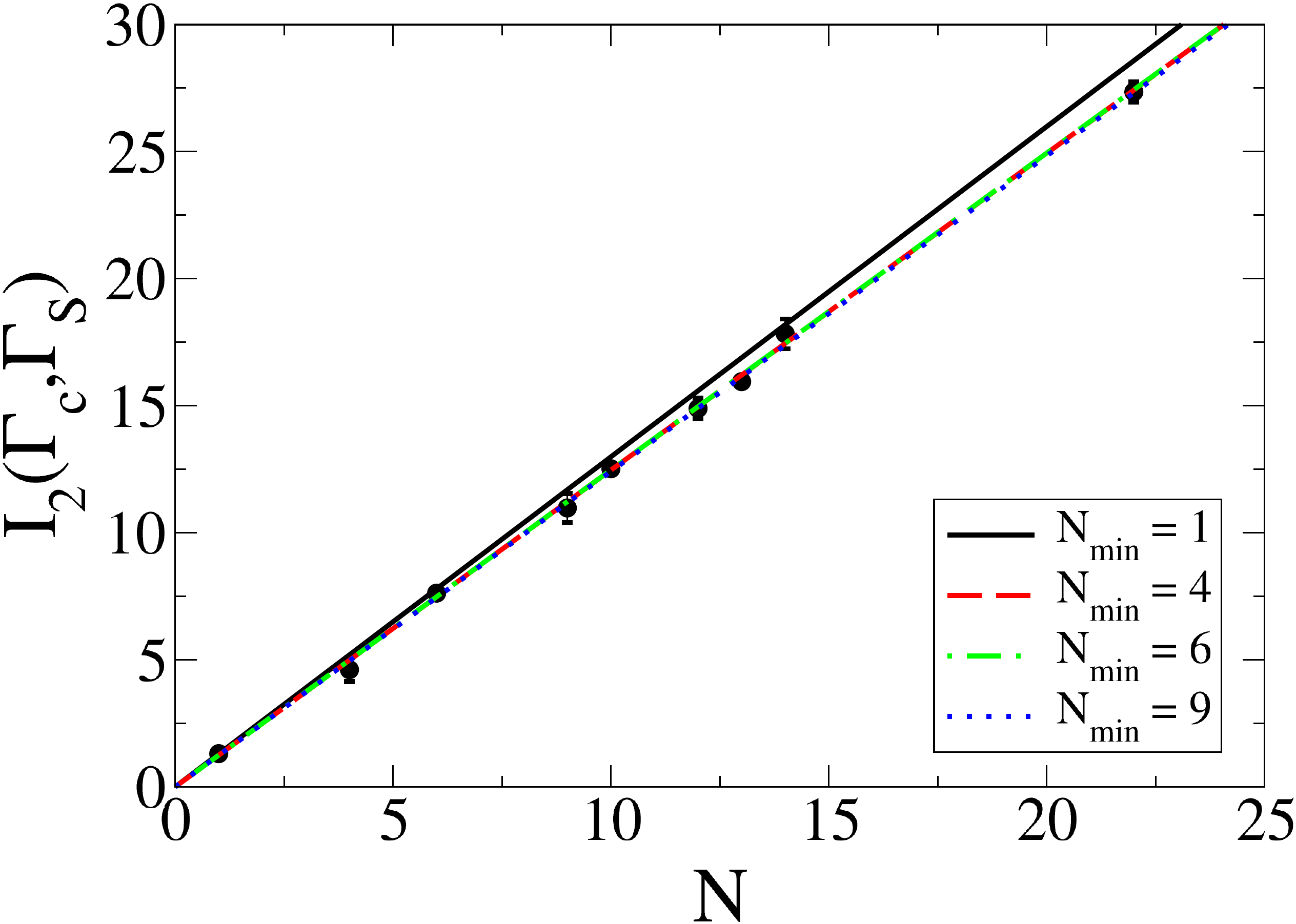}
  \caption{Mutual information $I_2(\Gamma_c, \Gamma_S)$ at $J_{\text{K}}=2.7$, $J_z=0$, $\beta=60$ for the choices of $\Gamma_c$, $\Gamma_S$ shown in Fig.~\ref{all_subsystems}, as a function of the size $N$ of $\Gamma_c$ and $\Gamma_S$. We compare with a linear fit to the right-hand side of Eq.~(\ref{arealaw}), for different choices of the minimum subsystem size $N_{\text{min}}$ taken into account.}
  \label{example_Jk2.7}
\end{figure}

\begin{figure}[b]
  \centering
  \includegraphics[width=0.85\linewidth]{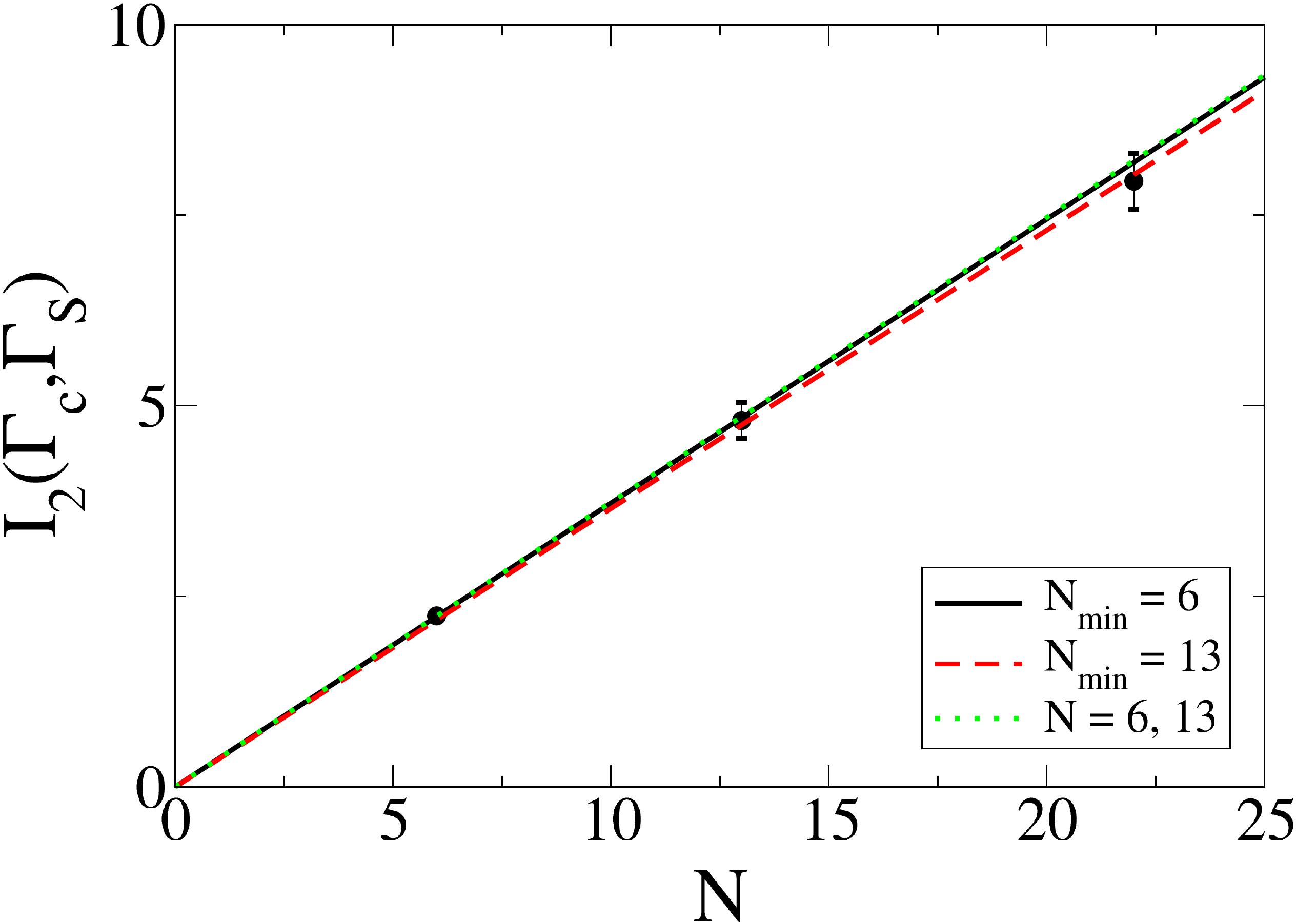}
  \caption{Same as Fig.~\ref{example_Jk2.7} for $J_{\text{K}}=1.8$, $J_z=1.8$, $\beta=40$, in the $z-$PKS phase, and for subsystems no. $3$, no. $7$, and no. $9$, shown in Fig.~\ref{all_subsystems}. We compare with three linear fits, obtained by using all data ($N_{\text{min}} = 6$), disregarding the smallest size ($N_{\text{min}} = 13$), and disregarding the largest size ($N=6,13$).}
  \label{example_Jk1.8_Jz1.8}
\end{figure}

\section{\uppercase{Computation of the mutual information}}
\label{sec:mutual}
In order to compute the coefficient $\alpha$ introduced in Eq.~(\ref{arealaw}) we have sampled $I_2(\Gamma_c, \Gamma_S)$ choosing different subsets of lattice sites. As explained in the main text, $\Gamma_c$ is a compact subset of lattice sites of the conduction electrons and $\Gamma_S$ contains the localized spin sites coupled to $\Gamma_c$. In Fig.~\ref{all_subsystems} we illustrate the nine choices of  $\Gamma_c$, $\Gamma_S$ used here. In the presence of nonvanishing antiferromagnetic coupling $J_z$, due to the enlarged unit cell \cite{SAG-18} we use only the subsystems no. $3$, no. $7$, and no. $9$, shown in Fig.~\ref{all_subsystems}.
As we discuss in the main text, for a given value of $J_{\text{K}}$ and $J_z$, the coefficient $\alpha$ has been obtained by a linear fit of $I_2(\Gamma_c, \Gamma_S)$ as a function of the size $N$. In Figs.~\ref{example_Jk2.7} and \ref{example_Jk1.8_Jz1.8} we show two examples of such a procedure, in the KI phase and in the $z$-PKS phase.

\section{\uppercase{Renyi entropies in the $z$-PKS phase}}
\label{sec:renyi}
\begin{figure}
  \centering
  \includegraphics[width=0.85\linewidth]{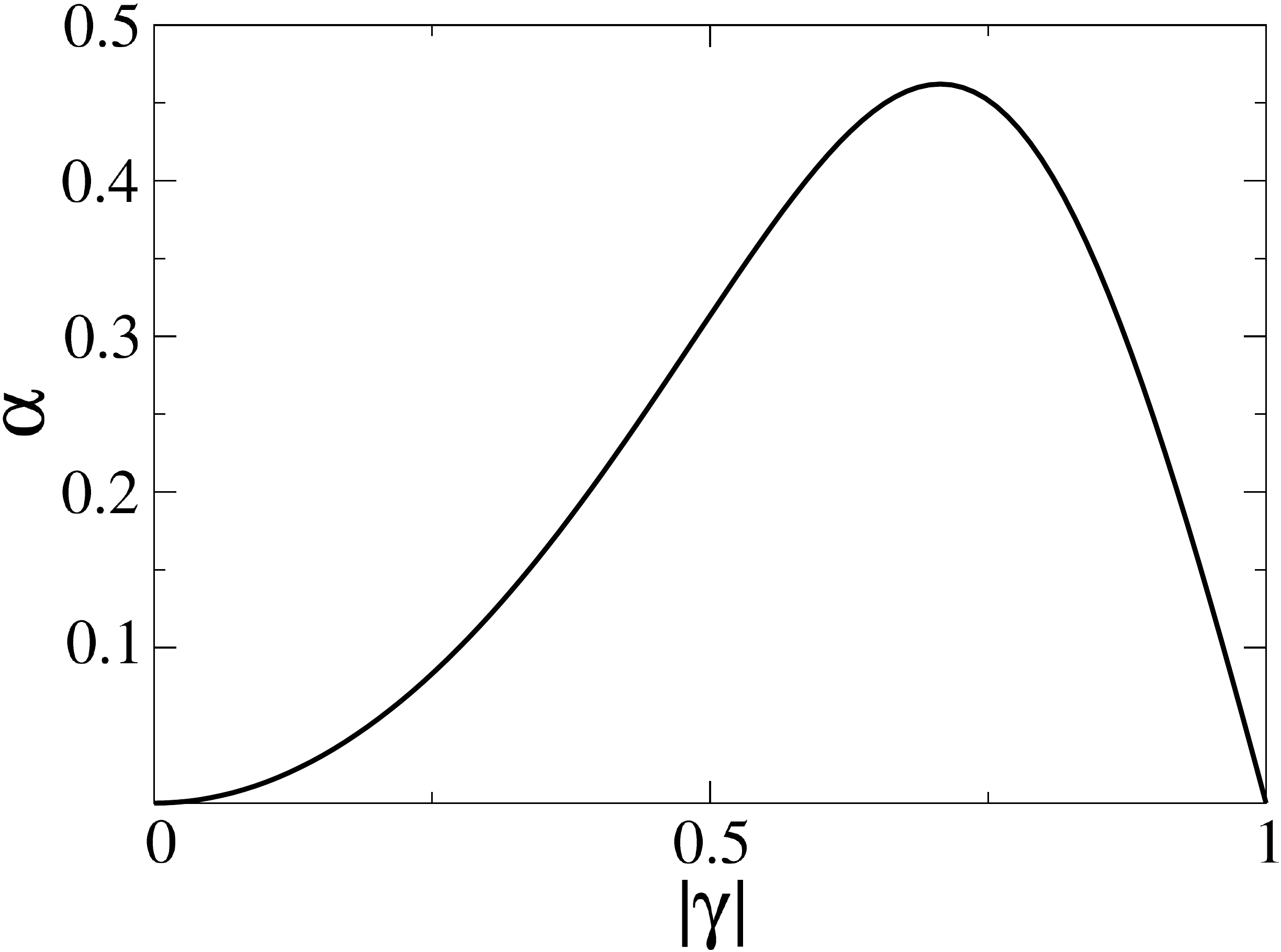}
  \caption{Coefficient $\alpha$ of the mutual information area law for the wave function ansatz of Eq.~(\ref{gs}).}
    \label{alpha_ansatz_fig}
\end{figure}

For a product wave function ansatz like that in Eq.~(\ref{gs}), and for any choice of $\Gamma_c$ and $\Gamma_S$, consisting in a set of $N$ conduction electron sites and in the set of the corresponding coupled localized spin sites, the reduced density matrix of $\Gamma_{c}  \cup \Gamma_{S}$ is a pure state. Hence, $S_n(\Gamma_{c}  \cup \Gamma_{S})=0$ and $S_n(\Gamma_{c})= S_n(\Gamma_{S})$, such that,
 in order to determine $I_2(\Gamma_c, \Gamma_S)$ is it sufficient to compute the entanglement entropy of the localized spins $S_n(\Gamma_{S})$.
 On each triangular unit cell, the ansatz of Eq.~(\ref{gs}) results in a factorized reduced density matrix for $\Gamma_{S}$,
 \begin{equation}
   \rho_S \equiv \Tr_{c} |\Psi\>\<\Psi| = \rho_{\ve{i}} \rho_{\ve{j}} \rho_{\ve{k}},
 \end{equation}
 with $|\Psi\>$ as given in Eq.~(\ref{gs}) and
 \begin{equation}
   \begin{split}
     &\rho_{\ve{i}} = 1,\\
     &\rho_{\ve{j}} = \rho_{\ve{k}} = \sum_{\sigma=\uparrow,\downarrow} \<\sigma|\psi(\gamma)\> \<\psi(\gamma)|\sigma\>, \\
     &|\psi(\gamma)\> \equiv \gamma |+\> - \sqrt{1-|\gamma|^2}|-\>.
   \end{split}
   \label{singlerhos}
 \end{equation}
 Since, apart from a site index, the single-site reduced density matrices $\rho_{\ve{j}}$ and $\rho_{\ve{k}}$ are identical, in Eq.~(\ref{singlerhos}) and in the following,
 with a slight abuse of notation, we have dropped the site indexes $y$, $z$ in $|\psi(\gamma)\>$.
 Using the definition $|+\> \equiv |\uparrow, \downarrow\>$ and $|-\>\equiv|\downarrow, \uparrow \>$, the reduced density matrix for the sites $\ve{j}$ and $\ve{k}$ is readily computed as
 \begin{equation}
   \rho_{\ve{j}}= \rho_{\ve{k}} =(1-|\gamma|^2) |\downarrow \>\<\downarrow | + |\gamma|^2|\uparrow \>\<\uparrow|.
   \label{rhoyz}
 \end{equation}
 Using Eqs.~(\ref{singlerhos}) and (\ref{rhoyz}) the second Renyi entropy for the single triangular unit cell is
 \begin{equation}
   S_2 = -2\ln \left(1 - 2|\gamma|^2 + 2|\gamma|^4\right).
   \label{S2ansatz}
 \end{equation}
 Finally, the coefficient $\alpha$ defined in Eq.~(\ref{arealaw}) corresponding to the Renyi entropy of Eq.~(\ref{S2ansatz}) is
  \begin{equation}
   \alpha = -\frac{2}{3}\ln \left(1 - 2|\gamma|^2 + 2|\gamma|^4\right).
   \label{alpha_ansatz}
 \end{equation}
  In Fig.~\ref{alpha_ansatz_fig} we plot $\alpha$ as a function of $|\gamma|$, for $|\gamma| \le 1$. The minimum value of $\alpha$ is $\alpha=0$, and corresponds to $\gamma=0$, $1$: for such values of $\gamma$ the state is a product of a conduction electrons and a localized spins states. Viceversa, a maximum value of $\alpha=2\ln(2)/3$ is found for $\gamma=1/\sqrt{2}$, corresponding to the maximally entangled spin singlet state.

\bibliographystyle{apsrev4-1_custom}
\bibliography{francesco,toshihiro,fassaad}

\end{document}